\documentclass{iopart}
\usepackage{iopams}

\newcommand{\C}{\mathbb C}
\newcommand{\HH}{\mathfrak{H}}
\newcommand{\N}{\mathcal{N}}

\begin{document}
\title[The Construction of Some Important Classes of Generalized CSs ...]
{Construction of Some Important Classes of Generalized Coherent
States: the Nonlinear Coherent States Method}
\author[R Roknizadeh and M K Tavassoly]{R. Roknizadeh, and M. K. Tavassoly}
\address{Department of Physics, University of Isfahan, Isfahan,
         Iran}
 \begin{abstract}
   Considering some important classes of generalized coherent
   states known in literature, we demonstrated that all of them can be created
   via conventional fashion, i.e. the "lowering operator eigen-state" and the "displacement
   operator" techniques using the {\it "nonlinear coherent states"}
   approach. As a result we obtained a {\it "unified method"}  to
   construct a large class of coherent states which already have been introduced by
   different prescriptions.
 \end{abstract}
 To be appeared in {\it J. Phys. A}:{\bf 37}(2004)
\eads{\mailto{rokni@sci.ui.ac.ir},
\mailto{mk.tavassoly@sci.ui.ac.ir}}
 \pacs{42.50.Dv}
 \maketitle
\section{{\bf Introduction}}\label{sec-intro}
   Coherent states (CSs) are venerable objects in physics with many
   applications in most of the fields of physics and mathematical
   physics [Klauder(1985) and Ali(2000)] from solid states physics
   to cosmology and occurred in the core of quantum optics.
   Along generalization of coherent states recently J-P Gazeau
   and J R Klauder proposed new coherent states(GK) for systems
   with discrete or continuous spectra, which parameterized by two
   real and continuous parameters $J\geq 0$ and $-\infty < \gamma < \infty$, denoted by
   $|J, \gamma \rangle$ [Gazeau(1999)]. They involve the
   representation of the discrete series $su(1, 1)$ Lie algebra,
   Coulomb-like spectrum, P\"oschl-Teller and the infinite square
   well potentials [Antoine(2001)].

   More recently J R Klauder, K A Penson and J-M Sixdeniers
   introduced another
   important classes  of generalized CSs through solutions of
   Stieltjes and Hausdorff moment problem [Klauder(2001)]. The constructed CSs in
   their paper which we will denote them by ${|z\rangle_{_{KPS}}}$ have these
   properties: i) {\it normalization} ii) {\it continuity in the
   label} and iii) they form an {\it overcomplete} set which allows
   a resolution of unity with a positive weight function.Then they
   have studied the statistical properties of these states by
   calculating Mandel parameter and also the geometry of these
   states by a metric factor, both analytically and numerically. At
   last, they used some estimations, to find the proper Hamiltonian
   for some sets of their CSs.

   The generalized CSs mentioned above, neither defined
   as eigestates of an annihilation operator, known as Barut-Girardello(BG) CSs
   [Barut(1971)], nor resulted from the action of a displacement
   operator on a reference state, say vacuum state, known frequently as
   Gilmore-Perelomov(GP) CSs [Gilmore(1972) and Perelomov(1972)].
   In other words the algebraic and symmetry
   considerations of the above two sets of states have not been clear, there.
   The same situation holds for the generalized CSs proposed by  K A Penson and A I
   Solomon we denote them by $|z, q\rangle_{_{PS}}$ [Penson(1999)].
   Due to these, our aim is to establish these subjects, and also to find a
   simpler way to relate the constructed CSs to an exact expression
   of the Hamiltonian, out of any estimation. In a sense our
   present paper may be considered along the completion of the
   works of Gazeau \etal [Gazeau(1999)], Klauder \etal [Klauder(2001)] and
   Penson \etal [Penson(1999)].

   Our procedure is based on the conjecture that these states may be
   studied  in the so-called {\it nonlinear}(NL) {\it coherent states}
   or{\it$f$-deformed coherent states} category [Manko(1997)].
   This is the main feature of our work. Nonlinear CSs
   attracted much attention in recent years, mostly because they exhibit
   nonclassical properties. As we know, up to now many quantum
   optical states such as $q$-deformed CS [Manko(1997)], negative
   binomial state [Wang(1999) and WangFu(1999)], photon added (and subtracted) CS
   [Sivakumar(2000), Sivakumar(1999) and Naderi(2004)], the center of mass motion
   of a trapped ion [Matos(1996)], some nonlinear phenomena such as
   a hypothetical {\it "frequency blue shift"} in high intensity
   photon beams [Manko(1995)] and recently after proposing
   $f$-bounded CS [Re'camier(2003)], the binomial state (or
   displaced exited CS) [Roknizadeh(2004)] have been considered as
   some sorts of nonlinear CSs.

   We attempt now to demonstrate that all sets of KPS coherent states
   including PS coherent state and the discrete
   series representations of the group $SU(1, 1)$
   (BG and GP coherent states) and $su(1, 1)$-Barut-Girardello CSs for
   Landau-level(LL) [Fakhri(2004)] can be classified in the nonlinear CSs with some
   special types of nonlinearity function $f(n)$, by which we may obtain the {\it "deformed
   annihilation and creation operators"}, {\it "generalized displacement
   operator"} and the {\it "dynamical Hamiltonian"} of the system.
   Based on these results it will be possible to reproduce
   all of the above CSs through the conventional
   fashion, i.e. by annihilation and displacement operator
   definition. In a general view the formalism presented in this paper
   provides a {\it unified} approach to construct all the employed
   CSs already introduced in different ways. In another direction as a matter of fact
   introducing the ladder operators related to the above CSs
   may be regarded as a first step in the process of generation of
   these states in some experimental realization schemes in quantum
   optics, when one intended to perform the
   interaction Hamiltonian describes formally the
   interaction between atoms and electromagnetic field.

   In addition to these, three interesting and new remarkable points,
   which seemingly that have not been pointed out so far, emerges from
   our studies.
   The first is that we can construct a new family of
   CSs (named the {\it dual}  set), other than the KPS and PS coherent
   states [Ali(2004) and Roy(2000)].
   The second is that the
   Hamiltonian proposed in [Manko(1997)] and others who cited
   him [see e.g.: Sivakumar(2000)] must be reformed, in view of the {\it
   action identity} requirement imposed on the generalized CSs of
   Klauder [Klauder(1998)]. We should quote here that recently some
   authors (for instance see A H El Kinani and M Daoud [ElKinani(2003)])
   have used normal ordering form (factorization) for their
   Hamiltonians. But they sometimes  made this suggestion for drastic
   simplification, without any deep physical bases [Speliotopoulos(2000)].
   Indeed they used the {\it supersymmetric quantum mechanics} (SUSQM) techniques
   as a {\it "mathematical tool"} to find the ladder operators for their Hamiltonians.
   Interestingly our formalism for solvable Hamiltonians gives an
   easier and clearer method to obtain these operators whenever
   necessary. We will pay more attention to this result in the
   conclusion.
   Thirdly, our results give us the opportunity to observe that for
   some sets of the KPS coherent states, the nonlinearity
   phenomena will be visible in {\it"low
   intensities of light"}, the fact that have been hidden, until we
   discovered the nonlinearity nature of them explicitly. This would be
   important experimentally, since as Man'ko \etal experienced, the
   nonlinear phenomena in $q$-oscillators can be detected only in
   high intensity photon beams [Manko(1995)]. Therefore if anyone can
   generate these particular sets of KPS coherent states by interaction of
   a field and atoms, it will be more easier to detect this phenomena.

   The plan of this paper is as follows: for the sake of
   completeness  we will bring a brief review on the nonlinear CSs,
   as Man'ko \etal introduced in section 2, following with a
   review on KPS and GK coherent states  in section 3. Then the relation
   between KPS, PS and BG coherent states of $su(1, 1)$ with nonlinear CSs
   will be obvious in section 4, and
   so the generators of the deformed oscillator algebra, displacement type operator, and the
   proper Hamiltonian in each set of the above coherent states will be
   find in sections 5 and 6. Based on our results we will
   introduce a vast class of new
   generalized CSs ({\it dual} family of KPS and PS coherent states).
   Then we discuss about the extension of the procedure to the
   GK coherent states in section 7, and finally we present our conclusions.
\section{{\bf Nonlinear coherent states}}\label{sec-nl}
    Nonlinear CSs were first introduced explicitly in [Matos(1996) and Manko(1997)]
    but before them it is implicitly defined by
    Shanta \etal [Shanta(1994)] in a compact form.
    This notion  attracted much attention in physical literature  in
    recent decade, especially because of their nonclassical
    properties in quantum optics. Man'ko \etal's approach is based on the two
    following postulates.

    The first is that the standard annihilation and
    creation operators deformed with an intensity dependent function
    $f(\hat{n})$ (which is an operator valued function), according to
    the relations:
  \begin{equation}\label{nonl-annih}
     A=af(\hat{n})=f(\hat{n}+1)a
  \end{equation}
  \begin{equation}\label{nonl-creat}
     A^\dag=f^{\dag}(\hat{n})a^\dag = a^{\dag} f^{\dag}(\hat{n}+1)
  \end{equation}
     with commutators between $A$ and $A^\dag$ as
  \begin{equation}\label{comut}
    [A,A^\dag]=(\hat{n}+1)f(\hat{n}+1)f^\dag(\hat{n}+1)-\hat{n}f^\dag(\hat{n})f(\hat{n})
  \end{equation}
    where $a$, $a^\dag$ and
    $\hat{n}=a^\dag  a$ are bosonic annihilation, creation and number
    operators, respectively. Ordinarily the phase of $f$ is irrelevant and
    one may choose $f$ to be real and nonnegative, i.e.
    $f^\dag(\hat{n})=f(\hat{n})$. But to keep general
    consideration, we take into account the phase dependence of
    $f(\hat{n})$ in general formalism given in this section.

    The second postulate is that the Hamiltonian of the deformed oscillator in analog to the
    harmonic oscillator is found to be
 \begin{equation}\label{hamilt}
   \hat{H}_{_M}=\frac{1}{2}(A A^\dag+A^\dag A)
 \end{equation}
   which by Eqs. (\ref{nonl-annih}) and (\ref{nonl-creat}) can be
   rewritten as
 \begin{equation}\label{hamilt-f}
   \hat{H}_{_M}=\frac{1}{2}\left((\hat{n}+1)f(\hat{n}+1)f^\dag(\hat{n}+1)+\hat{n}
   f^\dag(\hat{n})f(\hat{n})\right).
 \end{equation}
   where by index $M$ we want to denote the Hamiltonian as introduced by Man'ko \etal.
   The single mode nonlinear CSs obtained as eigen-state of the
   annihilation operator is as follows:
 \begin{equation}\label{nonl-cs}
   |z\rangle_{_{NL}}=\N_f(|z|^2)^{-1/2}\sum_{n=0}^{\infty}C_n z^n |n\rangle
 \end{equation}
   where the coefficients $C_n$ are given by
 \begin{equation}\label{Cn}
   C_n=\left(\sqrt{[nf^\dag(n)f(n)]!}\right)^{-1} \quad C_0=1 \quad
   [f(n)]!\doteq f(n)f(n-1)\cdots f(1)
 \end{equation}
   and the normalization constant is determined as
 \begin{equation}\label{normaliz}
   \N_f(|z|^2)=\sum_{n=0}^{\infty} {|C_n|^2 |z|^{2n}}.
 \end{equation}
   In order to have states belonging to the Fock space, it is
   required that $0< \N_f(|z|^2) < \infty$, which implies that
   $|z|\leq \lim_{n \mapsto \infty}n[f(n)]^2$.
   No further restrictions are then put on $f(n)$. Now with the
   help of Eqs. (\ref {nonl-cs}) and (\ref{Cn}) the function $f(n)$
   corresponding to any nonlinear CS is found to be
 \begin{equation}\label{findf}
   f(n)=\frac{C_{n-1}}{\sqrt n C_n}
 \end{equation}
   which plays the key rule in our present work.
   To recognize the nonlinearity of any CS we can use this simple and
   useful relation; by this we mean that if $C_n$'s for any CS are
   known, then $f(n)$ can be found from Eq. (\ref {findf}); when $f(n)=1$
   or at most be only a constant phase,
   we recover the original oscillator algebra, otherwise it is
   nonlinear.
\section{{\bf KPS and GK generalized coherent states}}\label{kps}
  Along generalization of CSs  J R Klauder \etal [Klauder(2001)] introduced the states
 \begin{equation}\label{kps-cs}
  {|z\rangle_{_{KPS}}}=\N(|z|^2)^{-1/2}\sum_{n=0}^{\infty}\frac{z^n}
  {\sqrt{\rho(n)}}|n\rangle
 \end{equation}
  where  $\rho(n)$ satisfies $\rho(0)=1$ and the normalization constant is determined as
 \begin{equation}\label{kps-norm}
  \N(|z|^2)=\sum_{n=0}^{\infty}\frac{|z|^{2n}}{\rho(n)}.
 \end{equation}
 Comparing Eqs. (\ref {nonl-cs}) with (\ref {kps-cs}) we obtain: $C(n)=[\rho(n)]^{-1/2}$
 which describes the relation between KPS and nonlinear CSs.
 Indeed when $C_n=1/\sqrt{n!}$ we obtain CCS, otherwise (for instance
 in general when $C_n=1/\sqrt{\rho(n)}$) as in (10), we are led to
 nonlinear CSs, this is what we want to show. We will demonstrate
 the equivalence between all sets of KPS
  coherent states and the nonlinear CSs in section 6 in detail.
  These states possess three conditions (i)-(iii) stated in the introduction, by
  appropriately selected functions $\rho(n)$. The third condition,
  which is the most difficult and at the same time the strongest
  requirements of any sets of CSs, were proved appreciatively by
  them, through Stieltjes and Hausdorff power-moment
  problem. Explicitly for each set of CSs ${|z\rangle_{_{KPS}}}$, they
  found the positive weight function $W(|z|^2)$ such that
 \begin{equation}\label{kps-res1}
    \int\int_{\C} d^2z {|z\rangle_{_{KPS}}} W(|z|^2)_{_{KPS}}\langle
    z|=\hat{I}=\sum_{n=0}^{\infty}|n\rangle \langle n|
 \end{equation}
   where $d^2 z=|z|d|z|d\theta$. Strictly speaking, evaluating the integral over
   $\theta$ in the LHS of Eq. (\ref {kps-res1}), setting $|z|^2\equiv
   x$ and simplify it, we arrive finally at:
 \begin{equation}\label{kps-res2}
   \int_0^R x^n\tilde{W}(x)dx=\rho(n)  \qquad n=0, 1, 2, ...   \qquad
   0<R\leq\infty
 \end{equation}
    where the positive weight functions $\tilde{W}(x)=\frac{\pi
    W(x)}{\N(x)}$ must be determined.
    Extending the values of $n \in \mathbb{N}$ in (\ref {kps-res2}) to $s \in
    \mathbb{C}$, it can be rewritten as:
 \begin{equation}\label{kps-res3}
    \int_0^\infty x^{s-1}\tilde{W}(x)dx=\rho(s-1)
     \end{equation}
    for $R=\infty$, which is known as Stieltjes moment problem and
 \begin{equation}\label{kps-res4}
    \int_0^\infty x^{s-1}H(R-x)\tilde{W}(x)dx=\rho(s-1)
 \end{equation}
    for $R < \infty$, which is known as Hausdorff moment problem,
    with $H(R-x)$ as the Heaviside function. The positive weight
    functions $\tilde{W}(x)$ can
    be obtained, through Mellin and inverse Mellin transform
    techniques, when $\rho(n)$ is choosed. This  automatically guarantees the
    condition of resolution of the identity for the KPS states according to (\ref {kps-res1}),
    (for a brief and nice discussion see: Klauder(2001)
    and references therein, e.g.: Marichev(1983) and
    Prudinkov(1998)).

    Adopting certain physical criteria rather than imposing selected
    mathematical requirements, Klauder and Gazeau  by
    reparametrizing the generalized CSs $|z\rangle$  in terms of a two independent
    parameters  $J$ and $\gamma$,
    introduced the generalized CSs $|J, \gamma \rangle$, known
    ordinarily  as Gazeau-Klauder(GK) coherent states in the physical
    literature [Klauder(1998) and Gazeau(1999)]. These are
    explicitly defined by the expansion
 \begin{equation}\label{kps-jgama}
   |J, \gamma \rangle=\N(J)^{-1/2}
   \sum_{n=0}^{\infty}\frac{J^{n/2}e^{-ie_n\gamma}}{\sqrt{
   \rho(n)}}|n\rangle
   \end{equation}
   where the normalization constant is given by
 \begin{equation}\label{kps-norm1}
   \N(J)=\sum_{n=0}^{\infty}\frac {J^n}{\rho(n)}
 \end{equation}
   and $\rho(n)$ is a positive weight factor with $\rho(0)\equiv
   1$ by convention and the domains of $J$ and $\gamma$ are such that $J\geq 0$ and
   $-\infty < \gamma < \infty$. These states required to satisfy the
   following properties: i) continuity of labeling: if
   $(J, \gamma)\rightarrow (J', \gamma')$ then,
   $\||J,\gamma \rangle -|J', \gamma'\rangle\|\rightarrow 0$, ii)
   resolution of the identity: $\hat{I} = \int |J, \gamma \rangle
   \langle J, \gamma |d \mu (J, \gamma)$ as usual and two extra
   properties: iii) temporal stability: $ \exp (-i \hat{H} t) |J,
   \gamma \rangle = |J,\gamma + \omega t \rangle$ and iv) the action
   identity: $ H=\langle J, \gamma |\hat{H}|J, \gamma \rangle=\omega
   J$, where $H$ and $\hat{H}$ are classical and quantum mechanical
   Hamiltonians of the system, respectively. It must be understood
   that the forth condition forced the generalized CSs to have the
   essential property: {\it "the most classical quantum states"},
   but now in the sense of {\it energy} of the dynamical system, in the
   same way that the canonical coherent state(CCS) is a quantum state which its
   position and momentum expectation values obey the classical
   orbits of harmonic oscillator in phase space.

   In Eq. (\ref{kps-jgama}) the kets ${|n\rangle}$
   are the eigen-vectors of the Hamiltonian $\hat{H}$,  with the
   eigen-energies $E_n$
 \begin{equation}\label{kps-14}
  \hat{H}|n\rangle=E_n|n \rangle \equiv \hbar \omega e_n |n
  \rangle \equiv e_n|n\rangle \quad \hbar \equiv 1
  \quad \omega\equiv 1 \quad  n=0,1, 2, ... .
 \end{equation}
   The action identity {\it uniquely} specified $\rho(n)$ in terms of
   the eigen-values of the Hamiltonian $\hat{H}$ with a discrete
   spectrum $0=e_0<e_1<e_2<...$
 \begin{equation}\label{kps-en}
   \rho(n)=\Pi_{k=1}^n e_k\equiv[e_n]!.
 \end{equation}
   As an example, for the shifted Hamiltonian of harmonic oscillator
   we have the CCS denoted by $|J, \gamma \rangle_{CCS}$:
 \begin{equation}\label{kpsCC-16}
   |J, \gamma \rangle_{_{CCS}} =
   e^{-{J/2}}\sum_{n=0}^{\infty}\frac{J^{n/2}e^{-in\gamma}}{\sqrt{n!}}
   |n\rangle.
 \end{equation}
   Eq. (\ref {kps-en}) obviously states that $\rho(n)$ is directly
   related to the spectrum of the dynamical system. So every
   Hamiltonian {\it uniquely} determined the associated CS, although the
   inverse is not true. This is because of the existence of the
   isospectral Hamiltonians in the context of quantum mechanics
   [Fern'andez(1994) and Roknizadeh(2004)].
\section{{\bf The relation between some classes of generalized coherent states and
   the nonlinear coherent states}}\label{sec-relat}
   To start with we demonstrate the relation between the KPS and
   NL coherent states. As we have mentioned already by comparing Eqs.
   (\ref {nonl-cs}) and (\ref{kps-cs}), the coefficients $C_n$ can
   be determined. Then inserting $C_n$ and $C_{n-1}$ into (\ref{findf}) yields:
 \begin{equation}\label{kps-20}
   f_{_{KPS}}(\hat{n})=\sqrt{\frac{\rho(\hat{n})}{\hat{n}\rho(\hat{n}-1)}}
 \end{equation}
   which provides simply a bridge
   between KPS and NL coherent states. As a special case when
   $\rho(n)=n!$, i.e. the canonical CS, we obtain $f(n)=1$.
   We will demonstrate the equivalence between all sets of KPS
   coherent states and the nonlinear CSs in section 6.
   From Eq. (\ref {kps-en}) $e_n$
   can easily be found as a function of $\rho(n)$:
 \begin{equation}\label{kps-21}
   e_n=\frac{\rho(n)}{\rho(n-1)}.
 \end{equation}
   Remembering that $e_n$s are the eigen-values of the Hamiltonian, it will be obvious
   that neither every $\rho(n)$ of KPS coherent states nor every
   $f(n)$ of nonlinear CSs are physically acceptable, when the dynamics
   of the system (Hamitonian) is specified. Using Eqs. (\ref
   {kps-20}) and (\ref {kps-21}) we get
 \begin{equation}\label{fn-en}
   f_{_{KPS}}(n)=\sqrt{\frac{e_n}{n}}  \qquad e_n=n(f_{_{KPS}}(n))^2.
 \end{equation}
   Now by using (\ref{kps-20}) we are able to find $f(\hat{n})$ for
   all sets of KPS coherent states and then  the deformed
   annihilation and creation operators $A=af(\hat{n})$ and
   $A^\dag=f^\dag(\hat{n})a^\dag$ may easily be obtained. Returning to
   the above descriptions, the operators $A$ and $A^\dag$ satisfy the
   following relations :
 \begin{equation}\label{Adag1}
   A|n\rangle =\sqrt{e_n} |n-1\rangle
 \end{equation}
 \begin{equation}\label{Adag2}
   A^\dag|n\rangle =\sqrt{e_{n+1}} |n+1\rangle
 \end{equation}
 \begin{equation}\label{AAdag3}
   [A, A^\dag]|n \rangle =( e_{n+1}- e_n) |n\rangle  \quad [A,
   \hat{n}]=A  \quad [A^\dag, \hat{n}] = -A^{\dag}.
 \end{equation}
   Also note that $A^{\dag}A |n\rangle=e_n|n\rangle$, not equal to $n|n\rangle$ in general.
   With these results in mind it would be obvious that Eq. (\ref{fn-en})
   is not consistent with the relations (\ref{hamilt})
   and (\ref{hamilt-f}) for the Hamiltonian. A closer look at Eq. (\ref{fn-en})
   which is a consequence of imposing the action
   identity on the Hamiltonian  of the system leads us to obtain
   a new form of the Hamiltonian for the nonlinear CSs as
 \begin{equation}\label{kps-23}
   \hat{H}=\hat{n}f^2(\hat{n})=A^\dag A.
 \end{equation}
   After all this form of the Hamiltonian  may be considered as
   {\it ''normal-ordered''} of the Man'ko \etal Hamiltonian $H_M$, introduced
   in Eq. (\ref{hamilt}),
 \begin{eqnarray}\label{kps-24}
   \hat{H}=\;\;:\hat{H}_M: = \frac{1}{2}\;\; :A^\dag A+AA^\dag:.
 \end{eqnarray}
   Therefore the associated Hamiltonian for the KPS coherent state
   can be written as
 \begin{equation}\label{kps-25}
   \hat{H}_{_{KPS}}=\hat{n}(f_{_{KPS}}(\hat{n}))^2=\frac{\rho(\hat{n})}{\rho(\hat{n}-1)}.
 \end{equation}
   Comparing Eqs. (\ref {hamilt}) and (\ref {hamilt-f}) with the
   Eqs. (\ref {fn-en}), (\ref {kps-23}) and (\ref {kps-24})
   implies that if we require that the KPS and nonlinear CSs to
   possess the action identity property, the associated Hamiltonian when
   expressed in terms of ladder operators must be reformed in
   normal-ordered form.

   In summary our considerations enable one to obtain $f$-deformed
   annihilation and creation operators as well as the Hamiltonian for all
   sets of ${|z\rangle_{_{KPS}}}$ discussed in [Klauder(2001)], after
   demonstrating that for each of them there exist a special nonlinearity
   function $f(\hat{n})$.
   Before closing this section, we bring two illustrative
   examples to show the ability of our method.
   \vspace{4 mm}\\
   {\em Example $4.1$ Penson-Solomon generalized CSs:}
   \vspace{2 mm}\\
   As an  example we
   can simply deduce the nonlinearity function for the generalized
   CSs introduced by Penson and Solomon(PS) [Penson(1999)]:
 \begin{equation}\label{ps1}
   |q, z\rangle_{_{PS}} = \N (q, |z|^2)^{-1/2} \sum_{n=0}^{\infty}
   \frac{q^{n(n-1)/2}}{\sqrt{n!}}z^n |n\rangle
 \end{equation}
   where $\N (q, |z|^2)^{-1/2}$ is a normalization function and
   $\varepsilon(q, 0)=1, 0\leq q \leq 1$.
   This definition is based on an entirely analytical prescription, in which
   the authors proposed the generalized exponential function obtained from the following differential equation:
 \begin{equation}\label{difeq}
  \frac {d \varepsilon(q, z)}{dz}=\varepsilon(q, qz) \Rightarrow
  \varepsilon(q, z)=\sum_{n=0}^\infty \frac {q^{n(n-1)/2}}{n!}z^n.
 \end{equation}
   So they have not used the ladder(or displacement) operator
   definition for their states [Penson(1999)].
   The nonlinearity function, the annihilation
   and creation operators evolve in these states can
   be easily obtained  by our formalism as
 \begin{equation}\label{ps2}
   f_{_{PS}}(\hat{n})= q ^ {1-\hat{n}} \qquad A=a q^{1-\hat{n}}
   \qquad A^{\dag}=q^{1-\hat{n}} a^{\dag}.
 \end{equation}
   Therefore our method enables one to reproduce them
   through solving  the eigen-value equation:
   $A|z, q\rangle_{_{PS}}=a q^{1-\hat{n}}|z, q\rangle_{_{PS}}= z |z, q\rangle_{_{PS}}$,
   in addition to introducing a $\hat{n}$ dependent Hamiltonian describing the dynamics of the
   system:
 \begin{equation}\label{psham}
   \hat{H}_{_{PS}}=\hat{n} q ^{2(1-\hat{n})}.
 \end{equation}
   \vspace{4 mm}
   {\it Example $4.2$ Barut-Girardello CSs for $su(1, 1)$ Lie algebra:}
   \vspace{2 mm}\\
    As second example we refer to the
   Barut-Girardello(BG) CSs, defined for the discrete series
   representations of the group $SU(1, 1)$ [Barut(1971)]. Our formalism
   provides an easier tool to relate the $SU(1, 1)$ CSs of BG to
   both the P\"oschl-Teller(PT) and infinite square well potentials.
   The BG states decomposed over the number-state bases as:
  \begin{equation}\label{BG1}
   |z, \kappa \rangle_{_{BG}} =
  \N(|z|^2)^{-1/2}\sum_{n=0}^{\infty}\frac{z^n}{\sqrt{{n! \Gamma
   (n+2\kappa)}}}|n\rangle
 \end{equation}
   where $\N(|z|^2)^{-1/2}$ is a normalization constant
   and the label $\kappa$ takes only the values $1, 3/2, 2, 5/2, ...$.
   Using (\ref{findf}) for these states we find
   \begin{equation}\label{BG11}
   f_{_{BG}}(\hat{n})=\sqrt {\hat{n}+2\kappa -1} \qquad
   \hat{H}_{_{BG}}=\hat{n}(\hat{n}+2\kappa-1) \quad n=0,1,2,... .
 \end{equation}
   So we have the eigen-value equation
 \begin{equation}\label{BG2}
   \hat{H}_{_{BG}}(\hat{n})|n\rangle=n(n+2\kappa-1)|n\rangle
 \end{equation}
   with $e_n = n(n+2\kappa-1)$; When $\kappa=3/2$ and
   $\kappa=\lambda+\eta$ ($\lambda$ and $\eta$ are two parameter
   characterize the PT potential: $\kappa=[(\lambda + \eta+1)/2]>3/2$)
   we obtain the infinite square well and PT potentials, respectively [Antoine(2001)].
   Therefore we have established that the dynamical group
   associated with these two potentials is the $SU(1, 1)$ group.
   Also if we take into account the action of
   $A=af_{_{BG}}(\hat{n}), A^{\dag}=f_{_{BG}}(\hat{n})a^{\dag}$ and $[A,
   A^{\dag}]$ on the states $|\kappa, n\rangle$ we obtain
 \begin{equation}\label{BG3}
    A|\kappa, n\rangle=\sqrt{n(n+2\kappa-1)}|\kappa, n-1\rangle
 \end{equation}
 \begin{equation}\label{BG4}
    A^{\dag}|\kappa, n\rangle= \sqrt{(n+2\kappa)(n+1)}|\kappa,
    n+1\rangle
 \end{equation}
 \begin{equation}\label{BG5}
    [A, A^{\dag}]|\kappa, n\rangle=(n+\kappa)|\kappa, n\rangle.
 \end{equation}
   We conclude that the generators of $su(1, 1)$ algebra $L_-,L_+,L_{12}$
   can be expressed in terms of the
   deformed annihilation and creation operators including their commutators
   such that
 \begin{eqnarray}\label{BG6}
  & & L_- \equiv \frac{1}{\sqrt 2}A=\frac{1}{\sqrt 2}a
    f_{_{BG}}(\hat{n})\nonumber\\
  & &  L_+ \equiv \frac{1}{\sqrt 2}A^{\dag}=\frac{1}{\sqrt 2}
    f_{_{BG}}(\hat{n})a^{\dag}\nonumber\\
   & & L_{12}\equiv \frac{1}{2}[A, A^{\dag}].
 \end{eqnarray}

   We note that our approach not only recover the results of
   [Antoine(2001)] in a simpler and clearer manner, but also it gives the
   explicit form of the operators ${L_-,L_+,L_{12}}$ as some
   {\it "intensity dependent"} operators in consistence with the
   Holstein-Primakoff single mode realization of the $su(1, 1)$
   Lie algebra [Gerry(1983)].
  \vspace{4 mm}\\
  {\it Example $4.3$  $su(1, 1)$-Barut-Girardello CSs for Landau levels:}
  \vspace{2 mm}\\
   As an example closer to physics, it is well-known that
   the Landau levels(LL) is directly related to quantum
   mechanical study of the motion of a charged and spinless particle on a flat
   plane in a constant magnetic field [Landau(1977) and Gazeau(2002)]. Recently it
   is realized two distinct symmetries corresponding to these
   states, namely $su(2)$ and $su(1,1)$ [Fakhri(2004)]. The author
   showed that the quantum states of the Landau problem corresponding to the
   motion of a spinless charged particle on a flat
   surface in a constant magnetic field $\beta/2$ along $z-$axis may
   be obtained as:
\begin{equation}\label{LL-Fock}
   |n, m\rangle=\frac {e^{i m \varphi}}{\sqrt{2 \pi}}
   (\frac{r}{2})^{\frac{2\alpha +1}{2}}
   e^{-\beta r^2/8}L_{n,m}^{(\alpha,
   \beta)}(\frac{r^2}{4})
\end{equation}
   where $0\leq \varphi \leq 2\pi$,
   $\alpha > -1, n\geq 0, 0 \leq m \leq n$ and $L_{n,m}^{(\alpha,
   \beta)}$ are the associated Laguerre functions.
   Constructing the Hilbert space spanned by $\HH :=\{|n, m\rangle\}_{n\geq 0, 0\leq m \leq
   n}$, there it is shown that the Barut-Girardello CSs(BGCSs) type associated to this
   system can be obtained as the following combination of the orthonormal basis:
 \begin{equation}\label{fakhri}
  |z\rangle_m=\frac{|z|^{(\alpha+m)/2}}{\sqrt
  {I_{\alpha+m}(2|z|)}}\sum_{n=m}^\infty \frac{z^{n-m}}
  {\sqrt{\Gamma(n-m+1)\Gamma(\alpha+n+1)}}|n, m\rangle
 \end{equation}
   where $I_{\alpha+m}(2|z|)$ is the modified Bessel function of
   the first kind [Watson(1995)]. The states in (\ref {fakhri}) derived with the help of
   the lowering generator of the $su(1, 1)$ Lie algebra, the action of which is defined by the relation:
 \begin{equation}\label{lower}
   K_-|n, m\rangle = \sqrt {(n+\alpha)(n-m)} |n-1, m\rangle
   \qquad K_-|m, m\rangle=0.
 \end{equation}
    A deep inspection to the states in (\ref {fakhri})
    in comparison to the states in (\ref {BG1}) shows a little
    difference, in view of the lower limit of summation sign in the
    former equation. But this situation is similar to
    the states known as photon-added CSs [Agarwal(1991)] in the
    sense that both of them are combinations of Fock space,
    with a cut-off in the summation from
    below. This common feature leads us to go on with the same
    procedure that has been done already to yield the
    nonlinearity function of the photon-added CSs(PACSs) in [Wang(1999) and Sivakumar(1999)].
    We define the deformed annihilation operator and the
    nonlinearity function similar to PACSs as
 \begin{equation}\label{defAf}
   A=f(\hat{n})a, \qquad f(n)=\frac{C_n}{\sqrt{n+1}C_{n+1}}
 \end{equation}
   where we have used the states (\ref{nonl-cs}) (replacing $|n\rangle$ by $|n, m\rangle$)
   as the eigen-states of the new annihilation operator defined in (\ref{defAf}).
   Upon these considerations we can calculate the nonlinearity
   function for the $su(1, 1)$-BGCSs related to LL as follows:
 \begin{equation}\label{}
   f_{_{LL}}(\hat{n})=\frac{(\hat{n}-m+1)(\hat{n}+\alpha+1)}{\sqrt{\hat{n}+1}}.
 \end{equation}
   Hence again the ladder operators corresponding to this system may
   be obtained easily.
   Besides this, we will observe in the next section that one can also find displacement
   operators associated with the above three examples, as well as
   KPS and GK coherent states.
\section{{\bf Introducing the generalized displacement operators}}
\label{sec-displce}
   After we recast the KPS, PS and $su(1, 1)$ coherent states
   as nonlinear CSs and found a nonlinearity
   function for each set of them, it is now possible to construct all of
   them through a displacement type operator
   formalism. We will do this in two distinct ways.\\

   {\bf I)} B Roy and P Roy gave a proposition
    and defined two new operators as follows [Roy (2000)]:
 \begin{equation}\label{kps-25}
   B=a\frac{1} {f(\hat{n})}  \qquad
   B^\dag=\frac{1}{f(\hat{n})}a^\dag.
 \end{equation}
    Before we proceed ahead to further clarify the problem,
    an interesting result may be given here.
    Choosing a special composition of the operators $A$ in (\ref{nonl-annih}) and
    $B^{\dag}$ in (\ref {kps-25}), we may observe that $B^{\dag}A|n \rangle=n|n \rangle
    =A^{\dag}B|n \rangle$. Also due to the following commutation
    relations:
 \begin{equation}\label{generator}
    [A, B^{\dag}]=I,\quad [A, B^{\dag}A]=A, \quad [B^{\dag},
    B^{\dag}A]=-B^{\dag}
 \end{equation}
    the generators $\{A, B^{\dag}, B^{\dag}A, I\}$ constitute the
    commutation relations of the Lie algebra $h_4$.
    The corresponding Lie group is the well-known
    Weyl-Heisenberg(W-H) group denoted by $H_4$.
    The same situation holds for the set of generators $\{B, A^{\dag}, A^{\dag}B,
    I\}$, using $A^\dag$ in (\ref{nonl-annih}) and $B$ in (\ref {kps-25}).

    Coming back again to the Roy and Roy formalism,
    the relations in (\ref{kps-25}) allow one to define two generalized displacement operators
 \begin{equation}\label{kps-26}
   D'(z)=\exp(zA^\dag-z^ *B)
 \end{equation}
 \begin{equation}\label{kps-27}
   D(z)=\exp(zB^\dag-z^* A).
 \end{equation}
     Noting that $D'(z)=D(-z)^\dag = [D(z)^{-1}]^\dag$,
     it may be realized that the {\it dual} pairs obtained generally from the actions of
     (\ref {kps-26}) and (\ref {kps-27}) on the vacuum state are
     the orbits of a projective {\it nonunitary} representations of the
     W-H group [Ali(2004)], so we named {\it displacement type} or {\it generalized
     displacement} operator.
   Therefore it is possible to construct two sets of CSs, the first is the
   old one introduced in [Klauder(2001)]:
 \begin{equation}\label{kps-28}
     D(z)|0\rangle \equiv |z\rangle_{_{KPS}}
 \end{equation}
   and the other one which is a new family of CSs, named {\it
   "dual states"} in [Roy(2000) and Ali(2004)] is:
 \begin{equation}\label{kps-29}
   D'(z)|0\rangle \equiv |z\rangle_{_{KPS}}^{dual}
   =\N(|z|^2)^{-1/2}\sum_{n=0}^{\infty}\frac{z^n
   \sqrt{\rho(n)}} {n!}|n\rangle
 \end{equation}
   where the normalization constant is determined as:
 \begin{equation}\label{kps-30}
   \N(|z|^2) = \sum_{n=0}^{\infty}\frac{|z|^{2n}\rho(n)}{(n!)^2}.
 \end{equation}
   Obviously the states $|z\rangle_{_{KPS}}^{dual}$ in (\ref {kps-29})
   are new ones, other than
   $|z\rangle_{_{KPS}}$. By the same procedures we have done in the
   previous sections it may be seen that the new states,
   $|z\rangle_{_{KPS}}^{dual}$ can also be considered as NL coherent
   state with the nonlinearity function
 \begin{equation}\label{kps-60}
   f^{dual}_{_{KPS}}(\hat{n})=\sqrt {\frac{\hat{n}
   \rho(\hat{n}-1)}{\rho(\hat{n})}}
 \end{equation}
   which is exactly the inverse of $f_{_{KPS}}(\hat{n})$, as one may
   expect. Also the Hamiltonian for the dual oscillator is found
   to be
 \begin{equation}\label{dual-H}
   \hat{H}^{dual}_{_{KPS}}
   =\hat{n}\left(f^{dual}_{_{KPS}}(\hat{n})\right)^2=\hat{n}^2
   \hat{H}_{_{KPS}}^{-1}.
 \end{equation}

   {\bf II)} More recently a general mathematical physics formalism
   for constructing the dual states has been proposed
   by S T Ali and us [Ali (2004)]. Following the latter
   formalism one can define the operator
 \begin{equation}\label{Top}
   \hat{T}=\sum_{n=0}^{\infty}\sqrt{\frac{n!}{\rho(n)}}|n \rangle \langle n|
 \end{equation}
   the action of which on canonical CS, $|z\rangle_{_{CCS}}
   =\exp{(-|z|^2/2)}\sum_{n=0}^{\infty}\frac{z^n}{\sqrt
   n!}|n\rangle$, yields the KPS coherent state:
 \begin{equation}\label{Top1}
   \hat{T} |z\rangle_{_{CCS}} = |z\rangle_{_{KPS}}.
 \end{equation}
   The $\hat{T}$ operator we introduced in Eq. (\ref {Top}) is well-defined and
   the inverse of it can be easily obtained as
 \begin{equation}\label{T}
   \hat{T}^{-1} = \sum_{n=0}^{\infty}\sqrt{\frac{\rho(n)}{n!}}|n \rangle \langle n|
 \end{equation}
   by which we may construct the new family of dual states:
   $\hat{T}^{-1} |z\rangle_{_{CCS}}  \equiv |z\rangle_{_{KPS}}^{dual}$.
   It is readily found that these states are just
   the states we have obtained in (\ref{kps-29}).
   Before applying the formalism to the KPS coherent states we
   give some examples.

   It must be mentioned that investigating the resolution of the identity and
   discussing the nonclassical properties, such as squeezing of the
   quadratures, sub (supper)-Poissonian statistics, amplitude
   squared squeezing, bunching (or antibunching) and the metric
   factor of these new states in (\ref{kps-29}) employing the same $\rho(n)$'s
   proposed in [Klauder(2001)] remain for future works.
   \vspace{4 mm}\\
   {\it Example $5.1$ Generalized displacement operator for PS
   coherent states:}
   \vspace{2 mm}\\
   Upon using the results we derived in example
   4.1 the dual of the PS states can be easily obtained using
   Roy and Roy approach [Roy(2000)]:
 \begin{equation}\label{ps5}
   |q, z\rangle^{dual}_{_{PS}} = \N (q, |z|^2)^{-1/2} \sum_{n=0}^{\infty}
   \frac{q^{-n(n-1)/2}}{\sqrt{n!}}z^n |n\rangle
 \end{equation}
   where $\N (q, |z|^2)$ is some  normalization constant, which may be
   determined. For this example the proposition in [Ali(2004)]
   works well. The $\hat{T}$-operator in this case reads:
 \begin{equation}\label{ps6}
   \hat{T}= \sum_{n=0}^{\infty}
   q^{n(n-1)/2} |n\rangle \langle n|
 \end{equation}
   by which we may obtain:
 \begin{equation}\label{ps7}
   \hat{T}^{-1}|z\rangle_{_{CCS}} \equiv \left(\sum_{n=0}^{\infty}
   q^{-n(n-1)/2} |n\rangle \langle n|\right)  |z\rangle_{_{CCS}}=|z, q\rangle_{_{PS}}^{dual}
 \end{equation}
    which is exactly the dual states we obtained in
    Eq. (\ref{ps5}).
   \vspace{4 mm}\\
   {\it Example $5.2$ Generalized displacement operators for BG and GP
   coherent states of $su(1, 1)$ Lie algebra:}
   \vspace{2 mm}\\
   As a well-known example we express the dual of
   BG coherent states in example 4.2.
   The duality of these states with the so-called
   Gilmore-Perelomov(GP) CSs have already been demonstrated in [Ali(2004)].
   The latter states were defined as:
 \begin{equation}\label{GP1}
   |z,\kappa\rangle_{_{GP}}=\N(|z|^2)^{-1/2}\sum_{n=0}^{\infty}\sqrt{\frac{(n+2\kappa-1)!}{n!}}
   z^n |n\rangle.
 \end{equation}
   where $\N (|z|^2)$ is a normalization constant.
   The nonlinearity function and the Hamiltonian may be written
   as: $f_{_{GP}}(\hat{n})=f_{_{BG}}^{-1}(\hat{n})$ and
   $\hat{H}=\hat{n}/(\hat{n}+2\kappa-1)$. Therefore we have in
   this case:
 \begin{equation}\label{GP2}
   B|\kappa, n\rangle=\sqrt{\frac{n}{n+2\kappa-1}}|\kappa,
   n-1\rangle
 \end{equation}
 \begin{equation}\label{GP3}
   B^{\dag}|\kappa, n\rangle=\sqrt{\frac{n+1}{n+2\kappa}}|\kappa,
   n+1\rangle
 \end{equation}
 \begin{equation}\label{GP4}
   [B, B^{\dag}]|\kappa, n\rangle=\frac{2\kappa-1}{(n+2\kappa)(n+2\kappa-1)}|\kappa,
   n\rangle.
 \end{equation}
   where $B=af_{_{GP}}(\hat{n})$ and $B^{\dag}=f_{_{GP}}(\hat{n})a^{\dag}$
   are the deformed annihilation and creation
   operators for the dual system (GP states), respectively. It is
   immediately observed that $[A, B^{\dag}]=I, [B, A^{\dag}]=I
   $, where obviously $A=af_{_{BG}}(\hat{n})$ and $A^{\dag}$ is its
   Hermition conjugate. So it is possible to obtain the
   displacement type operators for the BG and GP
   nonlinear CSs discussed in this paper, using relations (\ref{kps-26}) and
   (\ref{kps-27}). As a result the displacement operators
   obtained by our method are such that:
 \begin{equation}\label{GP6}
   |z,\kappa\rangle_{_{BG}}=D_{_{BG}}(z)|\kappa, 0\rangle=
   \exp{(zA^{\dag}-z^*B)}|\kappa, 0\rangle
 \end{equation}
    and
 \begin{equation}\label{GP7}
   |z,\kappa\rangle_{_{GP}}=D_{_{GP}}(z)|\kappa,
    0\rangle=\exp{(zB^{\dag}-z^*A)}|\kappa, 0\rangle.
 \end{equation}
   To the best of our knowledge these forms of displacement type operators for
   discrete series representation of $SU(1, 1)$ group, related to  BG and GP coherent
   states have not appeared in literature up to now.

   We close this section with implying that to apply the procedure to the $su(1, 1)$-BG
   coherent states for LL we expressed as an example (Example
   $4.3$), one must re-defined the auxiliary operators $B$ and
   $B^\dag$, in place of the ones introduced in (\ref {kps-25}) as follows:
\begin{equation}\label{BB-LL}
   B=\frac{1}{f_{_{LL}}(\hat{n})}a \qquad B^\dag =a^\dag
   \frac{1}{f_{_{LL}}(\hat{n})}.
\end{equation}
   The other calculations are the same as the general $su(1, 1)$
   coherent states in this example, so let us leave it over here.
 \section{{\bf Constructing KPS coherent states through annihilation and displacement operators
   and introducing proper Hamiltonians for them}}
 \label{sec-displce}
   Now we list, the nonlinearity function
   $f(\hat{n})$ and the associated Hamiltonian $\hat{H}(\hat{n})$, for
   all of the KPS  coherent states according to our proposed method.
   The operators $A$, $A^\dag$ can easily be obtained using Eqs.
   (\ref{nonl-annih}), (\ref{nonl-creat})
   and the displacement operators in like manner may be obtained
   using Eqs. (\ref{kps-25}), (\ref{kps-26}) and (\ref{kps-27}). So
   we ignore introducing them explicitly. A similar argument can also
   be made for the dual family of CSs (see (\ref {kps-60}) and
   (\ref{dual-H}) and the discussion before it for
   $f^{dual}_{_{KPS}}(\hat{n})$ and $\hat{H}^{dual}_{_{KPS}}(\hat{n}))$.
   Therefore introducing $f(\hat{n})$ and $\hat{H}(\hat{n})$, seems to be enough for
   our purpose.
   The set of CSs defined on the whole plane are as follows.
   \renewcommand{\labelenumi}{\alph{enumi})}
   \begin{enumerate}
 \item
   $\rho(n)=\frac{(n+p)!}{p!}$
 \begin{equation}\label{kps-da1}
   f(\hat{n})=\sqrt{\frac{\hat{n}+p}{\hat{n}}} \qquad  \hat{H}=\hat{n}+p.
  \end{equation}
   One can see from $\hat{H}$ that it is simply a shift in the energy
   eigen-states of the harmonic oscillator, as expected from
   $\rho(n)$, which is only a shift in $n!$.
   In all of the following cases we use the
   definitions  (\ref{Cn}) for $f(n)!$, whenever necessary.
 \item
   $\rho(n)=\frac{\Gamma(\alpha n+\beta)}{\Gamma (\beta)}$, where
   in this case and what follows $\Gamma$ is the gamma
   function. The CSs constructed by this $\rho(n)$ are known as the
   Mittag-Leffler (ML) CSs [Sixdeniers(1999)]. As a special case for
   $\alpha=1$ and $\beta$ arbitrary we obtain:
 \begin{equation}\label{kps-da2}
   f(\hat{n})=\sqrt{\frac{\hat{n}+\beta-1}{\hat{n}}} \qquad
   \hat{H}=\hat{n}+\beta-1
 \end{equation}
   where we have used the recurrence relation
   $\Gamma(z+1)=z\Gamma(z)$.  As a result we observe that the ML
   coherent states are also nonlinear CSs.
 \item
   $\rho(n)=\frac{n!}{n+1}$
 \begin{equation}\label{kps-da3}
   f(\hat{n})=\sqrt{\frac{\hat{n}}{\hat{n}+1}} \qquad
   \hat{H}= \frac{\hat{n}^2}{\hat{n}+1}.
 \end{equation}
   The dual of this state occurs for $\rho(n)=(n+1)!$, which is a
   special case of the item ($a$) when $p=1$. So there is no problem
   with the resolution of the identity for them.
 \item
 $\rho(n)=\frac{\Gamma(n+1+\alpha) }{\Gamma
   (1+\alpha)(1+n)}$
 \begin{equation}\label{kps-da4}
   f(\hat{n})=\sqrt{\frac{\hat{n}+\alpha}{\hat{n}+1}} \qquad
   \hat{H}=\frac{\hat{n}(\hat{n}+\alpha)}{\hat{n}+1}.
 \end{equation}
 \item
   $\rho(n)=(n!)^2$
 \begin{equation}\label{kps-da5}
   f(\hat{n})=\sqrt{\hat{n}} \qquad  \hat{H}= \hat{n}^2.
 \end{equation}
   This kind of deformation has already been employed by V Buzek
   [Buzek(1989)] who imposed on the single-mode field operators $a$
   and $a^\dag$, and showed that in the Jaynes-Cummings model with the
   intensity dependent coupling (when still the notion of nonlinear
   CS has not been used up to that time), interacting with the
   Holstein-Primakoff $SU(1, 1)$ CSs, the revivals of the radiation
   squeezing are strictly periodical for any value of initial
   squeezing. Also the dual of these states with $f(\hat{n})=1/\sqrt {\hat{n}}$
   has already been discovered and named {\it harmonious
   states} [Sudarshan(1993)].
 \item
   $\rho(n)=(n!)^3$
 \begin{equation}\label{kps-da6}
   f(\hat{n})=\hat{n} \qquad   \hat{H}= \hat{n}^3.
 \end{equation}
  \item
   $\rho(n)=\frac{n!\Gamma(n+3/4) }{\Gamma
   (4/3)}$
 \begin{equation}\label{kps-da7}
   f(\hat{n})=\sqrt{\hat{n}+1/3} \qquad  \hat{H}= \hat{n}(\hat{n}+1/3).
 \end{equation}
 \item
   $\rho(n)=\frac{(n!)^{3/2}\Gamma(3/2)}{\Gamma
   (n+3/2)}$
 \begin{equation}\label{kps-da8}
   f(\hat{n})=\frac{\hat{n}}{\sqrt{\hat{n}+1/2}} \qquad
   \hat{H}= \frac{\hat{n}^3}{\hat{n}+1/2}.
 \end{equation}
 \end{enumerate}
   All the Hamiltonians we derived above are such that
   $\lim_{n\rightarrow \infty} E_n  = \infty$. Upon more
   investigation in the above $f$- functions, an interesting physical point
   which may be explored is that the $f(n)$ functions in
   (\ref{kps-da1}), (\ref{kps-da2}), (\ref{kps-da3}) and
   (\ref{kps-da4}) will be equal to 1 (canonical CS) in the limit $n
   \rightarrow \infty$, i.e. for high intensities. So to observe the
   nonlinearity phenomena and its features in these special cases we do not need high
   intensities of light. On the contrary, if anyone can generate these
   special states through some physical processes, for instance the field-atom
   interaction in a cavity, the nonlinearity effects can be
   detected in {\it low intensities}.

   In addition to the above sets of CSs, Klauder \etal introduced a
   large class of CSs which have been defined on a unit disk. For these states
   also it is possible to continue in the same way we did for the
   states on the whole of the complex plane:
   \renewcommand{\labelenumi}{\alph{enumi}$^\prime$)}
\begin{enumerate}
\item
   $\rho(n)=\frac{2}{n+2}$
 \begin{equation}\label{kps-da}
   f(\hat{n})=\sqrt{\frac{\hat{n}+1}{\hat{n}(\hat{n}+2)}} \qquad
   \hat{H}=\frac{\hat{n}+1}{\hat{n}+2}.
 \end{equation}
 \item
   $\rho(n)=\frac{6}{(n+2)(n+3)}$
 \begin{equation}\label{kps-da}
   f(\hat{n})=\sqrt{\frac{\hat{n}+1}{\hat{n}(\hat{n}+3)}} \qquad
   \hat{H}=\frac{\hat{n}+1}{\hat{n}+3}.
 \end{equation}
   \item $\rho(n)=\frac{\pi}{4}\frac{(n!)^2 }{\Gamma^2
   (n+3/2)}$
 \begin{equation}\label{kps-da}
   f(\hat{n})=\frac{2\sqrt {\hat{n}}}{2\hat{n}+1} \qquad
   \hat{H}= \frac{4 \hat{n}^2}{(2\hat{n}+1)^2}.
 \end{equation}
  \item  $\rho(n)=\frac{3\pi}{8}\frac{n!(n+1)!}{\Gamma
   (n+3/2)\Gamma (n+5/2)}$
 \begin{equation}\label{kps-da}
   f(\hat{n})=2\sqrt{\frac{\hat{n}+1}{(2\hat{n}+1)(2\hat{n}+3)}} \qquad
   \hat{H}=\frac{4\hat{n}(\hat{n}+1)}{(2\hat{n}+1)(2\hat{n}+3)}.
 \end{equation}
  \item   $\rho(n)=\frac{\Gamma(1+c-a)\Gamma(1+c-b)\Gamma(n+1)\Gamma(n+1-a-b)}
   {\Gamma(1+c-a-b)\Gamma(n+1+c-a)\Gamma(n+1+c-b)}$, setting $a=b=1/2$ and
   $c=3/2$ we have
 \begin{equation}\label{kps-da}
   f(\hat{n})=\frac{\sqrt{\hat{n}+1/2}}{\hat{n}+1} \qquad
   \hat{H}=\frac{\hat{n}(\hat{n}+1/2)}{(\hat{n}+1)^2}.
 \end{equation}
   \item $\rho(n)=\frac{3 \Gamma (5/2)(n+1)!}{(n+3)\Gamma(n+5/2)}$
 \begin{equation}\label{kps-da}
   f(\hat{n})=\sqrt{\frac{\hat{n}^2+3\hat{n}+2}{\hat{n}(\hat{n}+3)(\hat{n}+3/2)}}
   \qquad \hat{H}= \frac{\hat{n}^2+3\hat{n}+2}{(\hat{n}+3)(\hat{n}+3/2)}.
 \end{equation}
 \end{enumerate}
   The common property of these states is that the limits of $f(n)$
   and $H$ as $n$ goes to infinity
   are $0$ and $1$, respectively.

   One can easily check that using the above $f$- functions (defined
   on the whole plane or restricted to the open unit disk) and
   solving the eigen-value equation $A|z\rangle = z |z\rangle$, or
   acting the displacement operator $D(z)$, obtained from Eq. ({\ref
   {kps-28}}) on the vacuum state $|0 \rangle$ by the well-known
   procedures, correctly lead to the $|z\rangle_{_{KPS}}$. The
   same argument may be followed for the dual family
   $|z\rangle_{_{KPS}}^{dual}$.
 \section{{\bf A discussion on constructing GK coherent states by
          annihilation and displacement operators techniques}}

   J R Klauder asked a question, "what is
   the physics involved in choosing the annihilation operator
   eigen-states"? Therefore he and J-P Gazeau re-defined the generalized CSs with the
   four requirements (i)-(iv) mentioned before in section 3
   [Klauder(1998 and 2001) and Gazeau(1999)], nowadays known as GK coherent
   states (equation (16)). To find out the nonlinearity nature of the GK coherent states
   one may apply the method we proposed, on these states.
   As a result we arrive at the following expression for $f_{_{GK}}(\gamma, \hat{n})$:
  \begin{equation}\label{GK-nonlinear}
    f_{_{GK}}(\gamma, \hat{n})=e^{i\gamma(\hat{e}_n-\hat{e}_{n-1})}\sqrt{\frac{\rho(\hat{n})}
    {\hat{n}\rho(\hat{n}-1)}}
  \end{equation}
   where we have choosed the notation $\hat{e}_n \equiv
   \frac{\rho(\hat{n})}{\rho(\hat{n}-1)}$ for simplicity.
   We demonstrated in the previous sections that
   this is the starting point to define the GK
   coherent states of any physical system as the (deformed)
   annihilation operator eigen-states
   and associate with them a displacement(type) operator.
   In addition to deducing the above results in detail, we applied the procedure on
   the known examples in the literature such as
   Coulomb-like spectrum [Gazeau(1999)], the P\"oschl-Teller and
   the infinite square well potentials [Antoine(2001)], formally.

    But, although we have introduced an expression for the nonlinearity
   function for the GK coherent states in equation (82), by which
   we can formally proceed further and get the explicit form of the
   requested results, upon a closer look at this relationship
   one can see that  $f_{_{GK}}(\gamma, \hat{n})$,
   is not a {\it well-defined operator valued function},
   in view of high careful mathematical considerations. It should be
   understood, that clearly this result is not due to illegality of the proposed approach.
   In fact the difficulty  in this special case might be expected
   naturally, because of the relaxing of the
   {\it  holomorphicity} [Odzijewicz(1998)] requirement in
   the definition of the GK coherent states [Gazeau(1999)].

   Nevertheless one may still use the presented formalism
   to find some special dual family associated with GK coherent states,
   the explicit form of which will not introduce here,
   until we find a well-defined form of them in future.
   Therefore the difficulties that already mentioned
   in [Ali(2004)] with the dual family of GK coherent states
   remain unsolved. We shall pay more attention to this matter in
   a forthcoming paper.\\
  \section {Concluding remarks}
    In summary we can classsify the obtained results as follows.
  \renewcommand{\labelenumi}{\Roman{enumi})}
  \begin{enumerate}
  \item
   We connected some important
   classes of CSs: KPS, PS and GK coherent states and the discrete
   series representation of $su(1, 1)$ Lie algebra using NL coherent
   states method successfully. Nonlinear CSs encompasses all these states as
   special cases which are distinguishable from each other via the nonlinearity function $f(n)$.
   So we have obtained a {\it "unified method"} to
   construct all of these CSs which already have been introduced by
   different prescriptions. Therefore this work
   may be considered in parallel to the previous efforts [Shanta(1994) and Ali(2004)]
   for unification of a large class of generalized CSs.
   Our re-construction of these states by the standard definitions,
   i.e. annihilation operator eigen-states and displacement operator
   techniques, with the preceding results enriches each set of
   the above classes of CSs in quantum
   optics, in the context of each other. So we enlarged the
   classification of nonlinear CSs with states having the main
   property of the generalized CSs (the resolution of  the identity),
   considerably.
 \item
   We did not discussed  about the dual states in detail, indeed we wanted only to show that
   our method can active the Roy and Roy approach [Roy(2000)] to
   produce the dual of any generalized CSs that can be classified
   in the nonlinear CSs. So introducing the dual family of generalized
   CSs coherent states has been done briefly, since it was out of the
   scope of our present work.
   \item
    The result we obtained in equation $(27)$ is in fact the factorization
   implied in the literature, while authors deal with solvable potentials
   and try to find ladder operators (see for instance Daoud (2002) and Elkinani (2003)).
   According to the latter approach the one-dimensional
   supersymmetric quantum mechanics(SUSQM) provides
   an algebraic tool to define ladder operators for some exactly solvable potentials.
   Therefore it may be understood that our method can be considered as an alternative
   formalism parallel to the well-known SUSQM techniques to reach this purpose more easily.
   Precisely speaking, for Hamiltonians having the properties mentioned in (18)  i.e.
   for some special solvable systems, one can factorize the Hamiltonian
   as we did in (27). This is interesting that these two distinct
   approaches terminate to a common point.
   To clarify more we add here that the equations (8)-(11) in [Daoud(2002)], which
   describe the {\it action} of creation, annihilation and their commutators
   on the related Fock space(where the author did not introduce
   the explicit form of the related operators), also can be derived with the help of our
   method, at least formally(see the discussion at section $7$).
   But since we are looking for not only their actions but also
   the explicit form of the related operators, in terms of standard ladder operators
   and number operator, we did not bring them in this contents due to the ill-definition of the
   nonlinearity function (equation 82) .\\
   \item
   As a postscript we mention another point.
   Based on the latter potentiality of such a naive approach
   we proposed to find the creation and annihilation operators,
   it may be realized as a comment that our formalism provides a
   straightforward framework for producing the {\it photon-added}
   [Agarwal(1991)] or more precisely {\it exited} CSs
   corresponding to exactly solvable potentials, such as P\"oschl-Teller and
   Morse potentials [Popov(2003) and Daoud(2002)].
   The procedure is really manifest. Once the eigen-values of the Hamiltonian were
   known, the Gazeau-Klauder CSs can be constructed using Eq. (\ref{kps-jgama}). The next step
   is to find the nonlinearity function through which one
   can perform ladder operators with the help of our procedure.
   Finally the iterative action of the creation operator on the
   GK coherent states readily gives the excited CSs. In this way
   this procedure can be applied in general to all sets of CSs
   discussed in the present work to create excited CSs of  KPS, PS and $su(1, 1)$ types,
   as well as the GK coherent states in a simple manner.
   \end{enumerate}

 \ack{ The authors would like to thank Dr. M. H. Naderi for useful
    discussions, also for reminding some references
    and from the referees for useful comments they suggested to improve the
    presentation. Finally in acknowledgement of fruitful discussions
    have been done with Prof. S Twareque Ali about luminosity of the subject of section 7
    we should grateful to him.}

\end{document}